\documentclass[prl,twocolumn,showpacs,superscriptaddress,amsmath,amssymb]{revtex4}

\newif\ifpdf
    \ifx\pdfoutput\undefined
    \pdffalse 
    \else
    \pdfoutput=1 
    \pdftrue
    \fi

    \ifpdf
    \usepackage[pdftex]{graphicx}
    \else
    \usepackage{graphicx}
    \fi

\begin{document}

\title{Molecular engineering of antiferromagnetic rings for quantum
computation}

\author{F. Troiani}
\email[Corresponding author: ]{troiani.filippo@unimore.it}
\affiliation{INFM - S$^{3}$ National Research Center on nanoStructures
and bioSystems at Surfaces and Dipartimento di Fisica, Universit\`a di
Modena e Reggio Emilia, via G. Campi 213/A, I-41100 Modena, Italy}
\author{A. Ghirri}
\affiliation{INFM - S$^{3}$ National Research Center on nanoStructures
and bioSystems at Surfaces and Dipartimento di Fisica, Universit\`a di
Modena e Reggio Emilia, via G. Campi 213/A, I-41100 Modena, Italy}
\author{M. Affronte}
\affiliation{INFM - S$^{3}$ National Research Center on nanoStructures
and bioSystems at Surfaces and Dipartimento di Fisica, Universit\`a di
Modena e Reggio Emilia, via G. Campi 213/A, I-41100 Modena, Italy}
\author{S. Carretta}
\affiliation{INFM and Dipartimento di Fisica, Universit\`a di Parma,
Parco Area delle Scienze, I-43100 Parma, Italy}
\author{P. Santini}
\affiliation{INFM and Dipartimento di Fisica, Universit\`a di Parma,
Parco Area delle Scienze, I-43100 Parma, Italy}
\author{G. Amoretti}
\affiliation{INFM and Dipartimento di Fisica, Universit\`a di Parma,
Parco Area delle Scienze, I-43100 Parma, Italy}
\author{S. Piligkos}
\affiliation{Department of Chemistry, University of Manchester, Oxford
Road, Manchester M139PL, United Kingdom}
\author{G. Timco}
\affiliation{Department of Chemistry, University of Manchester, Oxford
Road, Manchester M139PL, United Kingdom}
\author{R.E.P. Winpenny}
\affiliation{Department of Chemistry, University of Manchester, Oxford
Road, Manchester M139PL, United Kingdom}

\begin{abstract}
The substitution of one metal ion in a Cr-based molecular ring with 
dominant antiferromagnetic couplings allows to engineer its level structure 
and ground-state degeneracy. Here we characterize a Cr$_7$Ni molecular ring 
by means of low-temperature specific-heat and torque-magnetometry measurements,
 thus determining the microscopic parameters of the corresponding spin 
 Hamiltonian. The energy spectrum and the suppression of the leakage-inducing 
 $S$-mixing render the Cr$_7$Ni molecule a suitable candidate for the qubit 
 implementation, as further substantiated by our quantum-gate simulations.
 \end{abstract}

\date{\today}
\maketitle

\ifpdf
    \DeclareGraphicsExtensions{.pdf, ..jpg} \else
    \DeclareGraphicsExtensions{.eps, .jpg} \fi

Due to their relative decoupling from the environment and to the resulting
robustness, electron spins in solid-state systems are currently considered
among the most promising candidates for the storing and processing of
quantum information (QIP)~\cite{lmf}.
In this perspective, an increasing interest has recently been attracted by
a novel class of molecular magnets, including both ferromagnetic~\cite{ leuenberg} and antiferromagnetic~\cite{meier} systems.
In the latter case the quantum hardware is thought as a collection (e.g.,
a planar array) of coupled molecules, each corresponding to a different
qubit.
A major advantage with respect to analogous schemes based on single-spin
encodings would arise from the larger dimensions of the physical subsystem,
and on the resulting reduction of the spatial resolution which is required
for a selective addressing of each qubit by means of local magnetic
fields~\cite{lmf}.
A detailed, system-specific investigation is however mandatory in order
to verify the actual feasibility of this approach, and to suitably engineer
the intra- and inter-cluster interactions, the coupling between the computational and the environmental degrees of freedom, as well as the gating
strategy.

It is the purpose of the present Letter to argue the suitability
of Cr-based antiferromagnetic molecular rings for the qubit
implementation on the basis of a detailed theoretical and
experimental investigation of its wavefunctions and energy levels,
and of its simulated time evolution as induced by sequences of
pulsed magnetic fields. Molecular rings~\cite{FeRing} are
characterized by a cyclic shape and by a dominant
antiferromagnetic coupling between nearest neighbouring ions. In
the absence of applied fields and for even numbers of spin
centers, their energy spectrum typically consists in a singlet
ground state and in characteristic rotational
excitations~\cite{RotBand}. The recently demonstrated substitution
of a Cr$^{3+}$ ion with a divalent transition metal~\cite{Cr7M},
provides an extra spin to the otherwise fully compensated
molecule: this may in turn result in the formation of a ground
state doublet energetically separated from the higher energy
levels, i.e. in a suitable level structure for the qubit
implementation.

In the octanuclear heterometallic ring of our present concern, one of the
Cr$^{3+}$ (s=3/2) ions is substituted by a Ni$^{2+}$ (s=1) one.
The cyclic molecule, with a diameter $d \sim 1 $~nm, is characterized by
a planar arrangement;
we thus define $ \theta $ to be the angle between the static magnetic field
${\bf B}_0$ and $\hat{\bf z}$, the latter being perpendicular to the ring
plane.
The spin Hamiltonian corresponding to the single molecular magnet
reads~\cite{Carretta03}:
\begin{eqnarray}
{\cal H}=
\sum_{i=1}^8 J_{i}\, {\bf s}_{i} \cdot {\bf s}_{i+1} +
\sum_{i=1}^8 d_i\, [s_{z,i}^2-s_i (s_i +1)/3] + \nonumber \\
\sum_{i<j=1}^8 {\bf s}_{i}\cdot {\bf D}_{ij}\cdot {\bf s}_{j} +
\mu_B \sum_{i=1}^8 g_i{\bf B}\cdot {\bf s}_{i} ,
\label{eq1}
\end{eqnarray}

\noindent where the first term accounts for the dominant isotropic exchange
interaction, the second and third ones describe the anisotropic local
crystal-field and the intracluster dipole-dipole interaction, respectively;
isotropic $g$ factors are assumed for the last, Zeeman term.
The anisotropic part of ${\cal H}$ does not commute
with the squared total spin operator ${\bf S}^{2}$, and thus mixes subspaces
corresponding to different values of the total spin ({\it S-mixing}).
Due to its reduced symmetry, ${\cal H}$ can no longer be independently
diagonalized within each $(2S+1)$-dimensional block: an efficient solution scheme, based on an irreducible tensor operator formalism, has therefore
been developed (see Ref.~\cite{Carretta03} and references therein) and 
applied to the present case.

\begin{figure}[tbp]
\begin{center}
\includegraphics[width=\columnwidth,height=7cm]{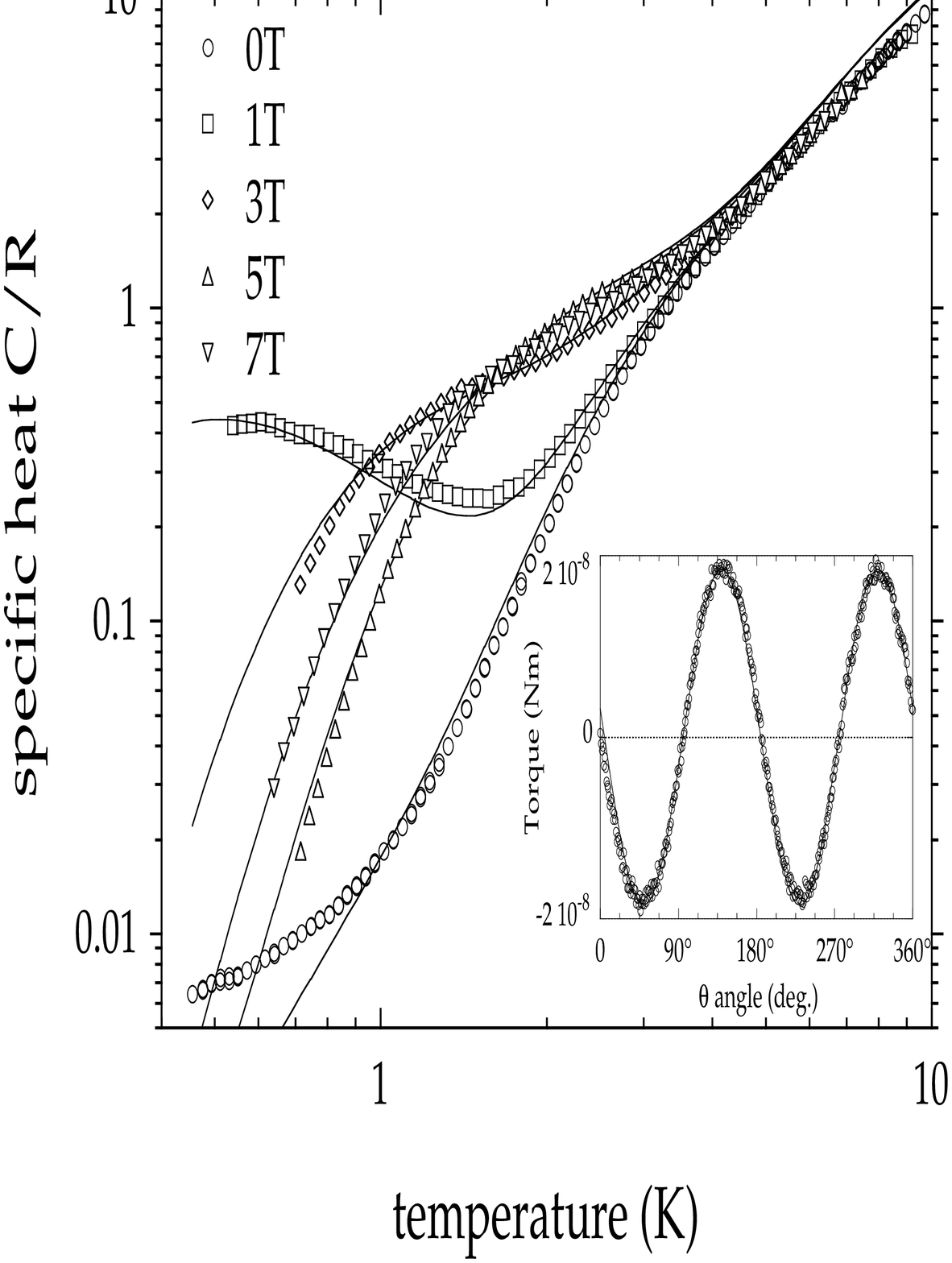}
\end{center}
\caption{Specific heat $ C / R $ of a single Cr$_7$Ni crystal
as a function of the temperature ($T$) and of the magnetic field ($B_0$).
We attribute the slight discrepancy between the theoretical curve and
the experimental data at $B_0=0$ and low $T$ to an experimental artifact, probably due to the presence of defected rings within the sample.
Inset: torque signal measured on a $ 0.7 \times 1 \times 0.5 $~mm$^3 $
Cr$_7$Ni crystal at $T=2$~K and $B_0=2$~T.
In our experimental set-up the rotation and torque direction is
perpendicular to ${\bf B}_0$ and a positive torque induces a
rotation of a negative $\theta$.}
\label{fig1}
\end{figure}

In order to estimate
the parameters entering the above spin Hamiltonian, we measure
the heat capacity $C$ as a function of the temperature ($ 0.4 < T < 10
$~K) and of the magnetic field ($ 0 < B_0 < 7 $~T)~\cite{Cr7M}.
The sample we investigate is a 2~mg bulk crystal, consisting in an ensemble
of independent, iso-oriented and identical rings.
In fact, the synthesis procedures developed in supramolecular chemistry
allow the engineering of regular arrays, as required
(together with an intercluster coupling) by the long-term goal of scaling
up the quantum hardware.
The experimental details and the method are described in Ref.~\cite{Carretta03}, 
while the
synthesis~\cite{Cr7M} and a systematic study of the thermodynamic
properties of [\{Me$_2$NH$_2$\}\{Cr$_7M$F$_8$(O$_2$CCMe$_3$)$_{16}$\}]
(in short Cr$_7$M, with M=Fe, Co, Ni, Mn, Cd) will be reported 
elsewhere.
In Fig.~\ref{fig1} we show $C(T;B_0)$, normalized to the gas constant $R$,
for five different values of $B_0$.
The typical low-$T$ Schottky anomaly appears upon application of the
magnetic field and is shifted towards higher temperatures as $B_0$
increases.
The overall specific heat $C(T;B_0)$ arises from two distinct contributions:
one is a lattice-related Debye term, $ C_{latt} / R = 234\, r\, T^3 /
(\Theta_D + \lambda T^2)^{3}$, where $r=298$ is the number of atoms per molecule,
$ \Theta_{D} = (158 \pm 10 ) $~K and $ \lambda = (0.42 \pm 0.02 )$~K$^{-1}$;
the other, $C_{m}$, depends on the spin degrees of freedom and is responsible for
the Schottky anomaly. The dependence of $C_m$ on the energy 
eigenvalues $\epsilon_i$ of $\mathcal{H}$ reads:
\begin{equation}
C_{m} = R \beta^2 \; \frac{\Sigma_i \epsilon_i^2 e^{-\beta \epsilon_i} \;
\Sigma_i e^{-\beta \epsilon_i} \; - \; (\Sigma_i \epsilon_i e^{-\beta \epsilon_i})^2}{(\Sigma_i e^{-\beta \epsilon_i})^2},
\label{eq2}
\end{equation}
\noindent with $\beta = 1 / k_BT$. The best fit of the
specific-heat temperature dependence at different values of $B_0$
(solid lines in Fig.~\ref{fig1}) provides the following set of
values for the microscopic physical parameters:  $ J_{Cr} / k_B = 
(17 \pm 0.5 )$~K, $ J_{Ni} / J_{Cr} = 0.9 \pm 0.1 $, $ | d_i | / k_B = 
(0.3 \pm 0.15) $~K
(for all the spin sites) and $g_{\rm Ni}=2.2\pm 0.1$; $g_{\rm Cr}=1.98 \pm 0.02$ 
was determined from previous measurements on Cr$_8$.
The sign of the anisotropies $d_i$
needs to be independently determined, as we do by means of
torque magnetometry measurements. In the inset of Fig.~\ref{fig1}
we report an angular scan of the torque signal, measured at
$T=2$~K and $B_0=2$~T. The torque exerted, e.g., for slightly
off-perpendicular magnetic-field orientations ($ \theta \sim
90^\circ $) tends to align the ring plane to the direction of
${\bf B}_0$. For symmetry reasons, the doublet ground state gives 
no contribution to the torque, and those of the first excited 
multiplet ($S=3/2$, see below) are the only excited states to be 
meaningfully populated at $T=2$~K: the above behaviour therefore
shows that the S=3/2 manifold has an easy-plane anisotropy, which
implies $d_i < 0$. Results independently obtained by high-field
torque measurements and inelastic neutron scattering (INS) confirm
the above estimates of the microscopic parameters~\cite{INS}. 
Lower-symmetry
terms probably need to be included in the spin Hamiltonian in
order to refine the interpretation of the INS data, but are
neglected in the following since, they have little relevance to
this discussion. Finally, the intracluster dipole-dipole interaction 
has been evaluated by means of the point-dipole approximation. 

We are now able to include the microscopic parameters in $
\mathcal{H} $ and accordingly draw the pattern of the low-lying
energy levels $\epsilon_i$ as a function of $ {\bf B}_0 = B_0
\hat{\bf z} $ (Fig.~\ref{fig2}). At zero field the ground state is
a degenerate doublet ($\epsilon_{0,1}$), with a largely dominating
($\agt 99\%$) total-spin component $S=1/2$; the first excited
states ($\epsilon_{2-5}$), instead, belong to a typical rotational
band, with $ S \simeq 3/2 $. Noticeably enough, these low values of 
the $S$-mixing allow us to consider the total spin as a good quantum 
number for the lowest eigenstates of the Cr$_7$Ni molecule.
As discussed in more detail in the following, two other physical
quantities also play a crucial role in the perspective of a QIP
implementation. The first one is the energy difference between the
ground-state doublet and the higher-lying levels, i.e. $\Delta
\equiv \epsilon_2 - \epsilon_1$: in fact, $\Delta$ determines to
which extent the ring behaves as an effective two-level system,
i.e. a meaningful population of any state but $ | 0 \rangle $ and
$ | 1 \rangle $ can be avoided throughout the molecule
manipulation. The second one is the splitting between the two
$S=1/2$ states, $\delta \equiv \epsilon_1 - \epsilon_0$: $\delta$
fixes the temperature the systems has to be cooled at in order for
it to be initialized to its ground state. 
Likewise the same existence of a ground-state doublet and the 
suppression of the $S$-mixing, the large energy separation 
from the higher states in the present molecule, $ \Delta (0) \simeq 
13$~K, is a non-trivial result of the system engineering. 
Besides, the magnetic field allows a further tuning of the molecule's
level structure. In particular, it increases
$\delta$, whereas it decreases the energy difference $\Delta
(B_0)$ between $ | 1 \rangle \simeq | S = 1/2, M = 1/2 \rangle $
and $ | 2 \rangle \simeq | 3/2, -3/2 \rangle $.
The achievement of the best trade-off between
the conflicting requirements of maximizing $\delta$ and $\Delta$
therefore determines the optimal value of the field, which we 
identify
\begin{figure}[tbp]
\begin{center}
\end{center}
\caption{Energy levels of the Cr$_7$Ni molecule as a function of a
static magnetic field applied along the $z$-axis. At zero field
the ground-state doublet is energetically separated from the
higher states ($\Delta (0) \simeq 13$~K), thus representing a
suitable choice for the qubit encoding.} \label{fig2}
\end{figure}
with $ B_0 = 2 $~T (see the discussion below). 
As a consequence, the achievable temperature required for
the system initialization to the $ | 0 \rangle $ state is $ T <<
\delta / k_B \simeq 2.4$~K, whereas $\Delta (B_0) \simeq \Delta
(0) -  2 g \mu_B B_0 \simeq 9.4$~K. On the grounds of the above
results, the Cr$_7$Ni molecule can be considered as an effective
$S=1/2$ spin cluster, and the information states $|0\rangle$ and
$|1\rangle$ safely identified with its ground-state doublet.
Such conclusion is unaffected by the uncertainty on $\Delta (0)$
(resulting from that on the microscopic parameters) which is roughly 
equal to that on $J_{Cr}$~\cite{error}.

The time simulation of the quantum gates provides an important
feedback for the optimization of the physical parameters. In fact,
the general unitary transformation applied to the computational
space is decomposed into a sequence of elementary gates, such as
the $SU(2)$ rotations of the single qubit and the two-qubit
$CNOT$~\cite{nielsen}. We start by considering the former, which
can be obtained as a combination of 3 rotations about any two
orthogonal axes, e.g. $ U (\alpha,\beta,\gamma) = \exp
(-i\alpha\sigma_2) \exp (-i\beta\sigma_1) \exp
(-i\gamma\sigma_2)$, being $\sigma_{1,2,3}$ the Pauli matrixes.
Transitions between the $|0\rangle$ and $|1\rangle$ states, 
i.e. rotations about the $x$ and $y$ axes, can be induced by means 
of resonant, in-plane electromagnetic pulses 
$ B_1 (t)\, \cos \omega t $, where
$B_1(t)\ll B_0$ represents the slowly-varying envelope. 
In the case of an effective two-level system, however, the transverse
magnetic field also couples the $ | 0 \rangle $ and $ | 1 \rangle
$ states to the higher-lying ones, thus inducing a population loss
({\it leakage}) during gating, quantified by $ L = 1 - [\,
|\langle 0\vert \psi(t)\rangle|^2 + |\langle 1\vert \psi(t)
\rangle |^2\,]$. More specifically, the occurrence of such
unwanted transitions is due to the $S$-mixing and to possible
intracluster inhomogeneities in the magnetic fields (or,
equivalently, in the ion $g$ factors): both result in
non-vanishing matrix elements $\langle 0,1 | {\cal H}_{op}| i \ge
2 \rangle $, with $\mathcal{H}_{op} = \mu_B \sum_{i=1}^8 g_i \,
{\bf B}_1 \cdot {\bf s}_{i}$. Together with a molecular
engineering aimed at the suppression of the $S$-mixing, the
minimization of $L$ can be achieved by the use of ``soft'' enough
pulses, i.e. by keeping the pulse spectral dispersion $ \Delta
\omega < (\Delta-\delta) / \hbar $. Being $ \Delta \omega \sim 1/
\tau_g $, such inequality gives a lower bound to the the ratio
$\tau_g/\tau_d$ that can in principle be achieved ($\tau_{g,d}$
are the gating and decoherence time, respectively).

Here, by means of a numerical integration of the Schr\"{o}dinger 
equation, we calculate the time dependence of $ | c_{0,1} |^2 = 
| \langle 0,1 | \psi (t) \rangle |^2 $ and of $L$ corresponding to 
a $\pi$ rotation about the $x$ axis, for $ | \psi (0) \rangle = | 0
\rangle $ and $ B_0 = 2 $~T (see Fig.~\ref{fig3}a); the pulsed
field is assumed to have a gaussian profile $ B_1 (t) = A \exp [-
( t - t_0 )^2 / ( 2 \sigma^2 ) ] $, with $A=0.1$~T and $\sigma =
150$~ps. To a very large extent the leakage involves the
first excited multiplet, and its value remains lower than $10^{-5}$ 
throughout the time evolution: the lower limit of $\tau_g$ imposed
by the presence of upper levels is thus by far larger than the  
one arising, e.g., from the present technological limitations in the pulse 
generation.
Further simulations, performed with more reasonable values  
of the magnetic field ($A=0.01$~T)~\cite{lidar}, give 
proportional increases of the pulse duration, witht a further 
suppression of $L$. We have also investigated the gate robustness with 
respect to possible spatial inhomogeneities of ${\bf B}_1$, generally 
leading to a larger values of $L$: even in the worst limiting case of 
a ${\bf B}_1$ which is nonzero only at the Ni-ion site, $L$ remains 
below the threshold of $10^{-4}$ ($A=0.15$~T, $\tau_g \sim 2\sigma \sim 
300$ps). 
A larger value of the static field $B_0$ would polarize nuclear spins, 
thus reducing the decoherence due to hyperfine field fluctuations (see 
below). 
For $B_0 > 5.6 T$, $\epsilon_1 > \epsilon_2$ and $\vert 1 \rangle$ 
can relax into $\vert 2 \rangle$ by emitting a phonon (the decay to 
$\vert 0\rangle$ is practically forbidden, owing to the approximate 
Kramers-doublet nature of the computational basis). 
These processes, however, are expected to be completely negligible with 
respect to other sources of decoherence, due to the small value of 
$(\epsilon_1 - \epsilon_2)/k_{B}\Theta_D$. 
An increase of $L$ might arise from the breakdown of the condition 
that $\Delta > \Delta \omega$. 
In order to estimate this effect, we have considered a static field 
$ B_0 = 5.61$~T, which implies $\Delta\simeq 0$ (Fig.~\ref{fig3}b): 
$L$ is seen to reach $\sim 3\%$, but even small deviations of $B_0$ 
from this critical values is enough to suppress $L$.
 
The experimental demonstration of the intercluster coupling required
for the implementation of the two-qubit gates is beyond the scope of 
the present work. In the following we discuss the present scenario 
and possible strategies to be pursued in order to achieve the required
conditional dynamics. 
In particular, links between Cr$_7$Ni rings formed by delocalized 
aromatic molecules have already been synthesized~\cite{timco},
and intercluster couplings have been demonstrated in similar systems~\cite{
coupling}. 
The simplest case occurs when the coupling between the rings results 
from the interaction of the $m$-th spin in molecule $I$ with the $n$-th 
of $II$.
If we apply the first-order perturbation theory and restrict ourselves to
the product space $\{ |0\rangle , |1\rangle \}_I \otimes \{ |0\rangle ,
|1\rangle \}_{II}$, a Heisenberg interaction $ \mathcal{H}_{I,II} = J^*
{\bf s}^I_m \cdot {\bf s}^{II}_n $ can be rephrased as follows in terms
of the cluster total spins:
\begin{equation}
\mathcal{H}_{I,II} = J_{eff} {\bf S}_I \cdot {\bf S}_{II},
\end{equation}
where
\begin{equation}
J_{eff} = 2/3 \, J^* \,
\langle 1/2 || s_m || 1/2 \rangle \,
\langle 1/2 || s_n || 1/2 \rangle 
\end{equation}
(analogous equations apply to the case of an Ising interaction).
In the present molecules the single-cluster reduced matrix elements $ \langle
1/2 || s_{m,n} || 1/2 \rangle $ has a modulus of the order of unity and
alternating positive-negative signs as a function of $n$ and $m$, whereas 
it is always negative for the Ni ion.
If the interchain couplings are more than one, $ J_{eff} $
is given by the sum of the single contributions.
Therefore, $J_{eff}$ has the same order of magnitude of the coupling $J^*$
between single spins, whereas its site dependence provides additional flexibility
to the implementation scheme.

\begin{figure}[tbp]
\begin{center}
\end{center}
\caption{Simulated time evolution of the molecular magnet, initially
prepared in $ | 0 \rangle $, under the effect of a transverse magnetic
field with a gaussian temporal profile. 
The black, red and blue lines correspond to $|c_0|^2$, $|c_1|^2$ and $L$
(multiplied by a factor $10^5$ in (a)), 
respectively; the static field is (a) $B_0=2$~T and (b) $B_0=5.61$~T, 
whereas $A=0.1$~T in both cases.}
\label{fig3}
\end{figure}

Within the present scenario the inter-cluster couplings, though
engineerable during the growth process, are untuneable thereafter.
This shortcoming is common to other implementations, and doesn't in
principle prevent from performing the single- and two-qubit quantum gates.
In fact, depending on the form and magnitude of the interchain coupling,
different approaches can be adopted.
Weak and Ising-like interactions allow the use of the so-called
``refocusing techniques'', widely developed within the NMR
community~\cite{nmr}.
Off-diagonal (e.g., Heisenberg) intercluster interactions
could instead require either multi-ring encodings of the qubit, resulting
in a symmetry-induced cancellation of the logical coupling between the
encoded qubits~\cite{zhou},
or the use of inter-qubit molecules acting as tuneable barriers~\cite{
benjamin}.

We finally discuss the decoherence of the cluster-spin degrees of freedom, 
which is expected to mainly arise from the hyperfine coupling with 
the nuclear spins.
A first estimate can be obtained by considering
the dipolar interaction of one Cr ion ($s$=3/2) with the neighboring F nucleus
(natural abundance $ \sim 100 \% $ of the $I=1/2$ isotope).
Being $ g_F = + 5.2577$ and the distance of each of the eight F atoms from
the nearest Cr ion $ d = 0.191 $nm, the interaction energy corresponds to
$ E_{hyp} / k_B = 0.38 $~mK. For an octanuclear ring this would give $ \tau_{d} \simeq \hbar /8 E_{hyp}
\sim 2.5$~ns, that can be considered as a lower bond for $ \tau_{d}$. Similar $ \tau_{d}$ values have 
been estimated for other molecular magnets \cite{coupling, delBarco}. 
Direct measurements of the electron-spin resonance linewidth on a 
Cr$_7$Ni crystal (which also includes the effects of possible inhomogeneities,
the so-called ``dephasing'') provides $ \tau_{d}$ values one order of magnitude larger 
than the above one ~\cite{WW}.
Substantial enhancements of $ \tau_d $ will result from the suppression
of the hyperfine field fluctuations (mK temperatures and few-T static
fields) and from the substituting of the F ions with OH groups in the
Cr$_7$Ni compound, which we are already working at.
Besides, the static field induces a large mismatch between the energy gaps 
of the nuclear and electronic spins, thus rendering highly inefficient the 
relaxation processes of the latter.
A second potential source of decoherence, namely the ring-ring dipolar
coupling, is characterized by a lower energy scale, $E_{dip}/k_B \simeq (g
\mu_B S)^2/V \sim 0.1 $~mK ($ S = 1/2 $ and $ V = 6.346 {\rm nm}^3 $), futher
reduceable in diluted Cr$_7$Ni molecular systems.

In conclusion, the energy spectrum of the investigated Cr$_7$Ni molecule 
fully justifies its description in terms of an effective two-level system;
besides, the symmetries of the ground-state doublet ($S$-mixing below 1\%)
suppress the coupling to the higher levels as induced by the transverse 
magnetic fields which are required for the quantum-gate implementation.
In fact, our simulations of the single-qubit gates provide negligible 
values for the leakage ($L\alt 10^{-4}$) even for gating times of the 
order of $10^2$~ps, i.e. well below the tens of ns estimated for the spin 
decoherence times. While further work is needed for the engineering of the
intercluster coupling, these results strongly support the suitability of 
the Cr$_7$Ni rings for the qubit encoding and manipulation.

\end{document}